\documentclass[pra,twocolumn, amsmath,amssymb,superscriptaddress]{revtex4-1}
\usepackage[dvipdfmx]{graphicx}
\usepackage{dcolumn}
\usepackage{bm}
\usepackage{subfigure}

\def\r{\rangle}
\def\l{\langle}
\def\disp{\displaystyle}

\begin{document}

\title{Memory effect and pseudomode amplitude in non-Markovian dynamics of a two level system}

\author{Yuta Ohyama} 
\affiliation{Graduate School of Pure and Applied Sciences, University of Tsukuba, 1-1-1 Tennodai, Tsukuba, Ibaraki 305-8571, Japan}

\author{Yasuhiro Tokura}
\affiliation{Graduate School of Pure and Applied Sciences, University of Tsukuba, 1-1-1 Tennodai, Tsukuba, Ibaraki 305-8571, Japan}
\affiliation{NTT Basic Research Laboratories, NTT Corporation, 3-1 Morinosato Wakamiya, Atsugi, Kanagawa 243-0198, Japan}

\date{\today}

\begin{abstract}
We study non-Markovian dynamics of a two level atom using pseudomode method. 
Because of the memory effect of non-Markovian dynamics, the atom receives back information and excited energy from the reservoir at a later time, which causes more complicated behaviors than Markovian dynamics.
With pseudomode method, non-Markovian dynamics of the atom can be mapped into Markovian dynamics of the atom and pseudomode.  
We show that by using pseudomode method and quantum jump approach for Markovian dynamics, we get a physically intuitive insight into the memory effect of non-Markovian dynamics.
It suggests a simple physical meaning of the memory time of a non-Markovian reservoir.
\end{abstract}

\maketitle

\section{Introduction\protect\\}
All realistic quantum systems are open quantum system; the system interacts with reservoir systems which cause decoherence and relaxation \cite{breuer2007theory}.
According to characters of the interaction and the structure of the reservoirs, the dynamics of open quantum systems can be classified into Markovian dynamics with no memory effect and non-Markovian one with memory effect.

In Markovian open system, the reservoir acts as a sink for the system information; the information that the system of interest loses into the reservoir does not play any further role in the system dynamics.
However, in non-Markovian case, this lost information is temporarily stored in the reservoir and comes back at a later time to influence the system \cite{RevModPhys.88.021002}.
This is the memory effect of non-Markovian dynamics and causes more complicated behaviors than Markovian dynamics.

There are stochastic approach for Markovian dynamics \cite{RevModPhys.70.101,Carmichael1993Open,wiseman2010quantum,PhysRevLett.68.580,1355-5111-8-1-007,PhysRevA.47.1652}; quantum jump, Monte Carlo wave function and quantum trajectory.
In these methods, the dissipation caused by the interaction with the reservoir is interpreted as an incoherent jump between two states and the state of the system is described by the sum of the ensembles which are identified by the jump times.
Therefore, we can get the intuitive understanding of the dynamics.

Recently, non-Markovian dynamics has been investigated \cite{PhysRevLett.103.210401,PhysRevA.81.062115,PhysRevA.81.062124}.
In these papers, non-Markovianity of quantum processes is discussed.
The measures for the degree of non-Markovianity are based on the distinguishability of quantum states,
which focuses on the dynamics of the system of interest.
In this paper, we use pseudomode method \cite{PhysRevA.54.3592,PhysRevA.55.2290,PhysRevA.55.4636,PhysRevA.80.012104,PhysRevA.90.054101,PhysRevA.90.062104}.
With pseudomode method, non-Markovian dynamics of the system of interest can be mapped into Markovian dynamics of a combined system of the system of interest and pseudomode.
Therefore, the dynamics of the extended system can be discussed.
Non-Markovian quantum jump has been also investigated \cite{PhysRevLett.100.180402,PhysRevA.79.062112,PhysRevA.86.022102}.
Because of the memory effect and back flow from the reservoir into the system, the description is more complicated than Markovian case and  pure state quantum trajectories for general non-Markovian systems do not exist \cite{PhysRevLett.101.140401}.
By connecting this method and pseudomode method, we get  a simple intuitive physical picture of the memory of a non-Markovian reservoir and of how such memory allows to partly restore some of the coherence lost to the environment \cite{PhysRevA.80.012104}.
This result also suggests that pseudomode could be seen as an effective description of the reservoir memory.

The purpose of this paper is to get a more physically intuitive insight into the memory effect of non-Markovian dynamics.
For this purpose, we use pseudomode method and quantum jump approach.
With pseudomode method, non-Markovian dynamics of the system is described by Markovian dynamics of a combined system, so that we can apply quantum jump approach for Markovian dynamics to the combined system.
The result gives us a simple physical meaning of the memory time of a non-Markovian reservoir.

The paper is organized as follows.
In Sec.~\ref{Model}, we persent the model discussed in the paper.
In Sec.~\ref{SA}, the property of the model we have presented in Sec.~\ref{Model} is evaluated using quantum jump approach for Markovian dynamics and, in  Sec.~\ref{DJC}, we study the dynamics of the damped Jaynes Cummings model, which is a typical example of non-Markovian system.
Finally, we conclude the paper in Sec.~\ref{conclude}.

\section{Model \protect\\ }\label{Model}
Non-Markovian systems appear in many branches of physics.
Here we consider a two level atom interacting with a structured electromagnetic reservoir which is described by the Jaynes-Cummings model with rotating approximation \cite{breuer2007theory}.
The Hamiltonian for the total system is 
\begin{align}
H=\dfrac{\hbar\omega_0}{2}\sigma_z +\disp\sum_k \hbar \omega_k b_k^\dag b_k+\disp\sum_k \hbar g_k \left(\sigma_+ b_k +\sigma_-b_k^\dag\right),
\end{align}
where $\sigma_z = |e\r\l e| - |g\r \l g|, \sigma_+ = (\sigma_-)^\dag = |e\r \l g|$, 
$b_k^\dag $ and $b_k$ are the bosonic creation and annihilation operators for the reservoir mode $k$ with frequency $\omega_k \geq 0$
and $g_k$ is the coupling between the two level atom and the reservoir mode $k$ in the reservoir.
$|e\r$ and $|g\r$ are excited and the ground states of the two level atom, respectively.
Total excitation number is a conserved quantity in this model.

Let the atom be in an arbitrary superposition and the reservoir be in vacuum state at $t=0$,
therefore the initial state is given by
\begin{align}
|\Psi(0)\r = (\alpha|e\r + \beta |g\r)\otimes |0\r,
\label{Psi_0}
\end{align}
where $|0\r$ denotes the vacuum state of the reservoir,
and $\alpha$ and $\beta$ satisfy the normalization condition $|\alpha |^2 + |\beta|^2=1$.
In the interaction picture, the total state at $t\geq 0$ can be expanded as
\begin{align}
|\Psi(t)\r_I = \alpha\left(a_{0}(t)|e,0\r+\sum_k a_{k}(t)|g,1_k\r \right)+\beta|g,0\r.
\label{Psi}
\end{align}
where $|1_k\r = b_k^\dag |0\r$ and the coefficient of $|g,0\r$ is independent of time.
Inserting this state into  Schr\"{o}dinger equation, we get the integro-differential equation for atomic amplitude $a_0(t)$,
\begin{align}
 \dfrac{d}{dt}a_{0}(t)= -  \int_0^t dt' f(t-t') a_{0}(t'), 
  \label{a_0}
\end{align}
where $f(t) \equiv \sum_k g_k^2 e^{-i (\omega_k-\omega_0)t} $ is a correlation function.

Here we assume that the coupling $g_k$ depends only on the frequency $\omega_k$.
In a continuous distribution limit, the sum on the reservoir mode $k$ is replaced by an integral by $\omega$ as follows,
\begin{align}
\sum_k g_k^2  \simeq \int_{-\infty}^\infty d\omega  \rho(\omega) g^2(\omega) = \dfrac{1}{2\pi}\int_{-\infty}^\infty d\omega  D(\omega),
\end{align}
where $\rho(\omega)$ is the density of states of the reservoir.
The structure of the reservoir is characterized by the positive definite function $D(\omega)$.
Because we have extended the integral to $-\infty$, this function should be vanished in the negative $\omega$ region.
With these equations, the correlation function becomes
 \begin{align}
 f(t) = \dfrac{1}{2\pi}\int_{-\infty}^\infty d\omega  D(\omega)e^{-i (\omega-\omega_0)t}.
 \end{align}

If the reservoir has no structure, $D(\omega) $ does not depend on $\omega$.
In this case, the correlation function is proportional to the delta function, so that the dynamics of the system is Markovian dynamics.
In the following, we restrict that $D(\omega)$ can be approximated by a sum of Lorentz functions.
This is not the necessary condition for using pseudomode method but an assumption for simplicity.
We set the explicit form of $D(\omega)$ as 
\begin{align}
D(\omega) \simeq \sum_{l =1}^L \dfrac{\gamma_l \lambda_l^2}{(\omega -\omega_l)^2 + \lambda_l^2},
\label{D_omega}
\end{align}
where $\gamma_l$ is the coupling strength and $\lambda_l^{-1}$ is the reservoir's correlation time. 
Since $D(\omega)$ is vanished in the negative $\omega$ region, the resonant frequency $\omega_l$ should be much lager than the width $\lambda_l$.

From the residue theorem and $t-t'\geq 0$, only the poles in the lower half plane have contribution for the dynamics of $a_0(t)$.  
So we define that $\lambda_l$ is positive for any $l$.
Since $D(\omega)$ should be non-negative for any $\omega$, we consider $\gamma_l >0$ for any $l$, which is also not the necessary condition but  an assumption for simplicity.
Integral of Lorentz function is
\begin{align}
\int_{-\infty}^\infty d\omega \dfrac{\lambda_l^2 }{(\omega -\omega_l)^2 + \lambda_l^2} =\pi \lambda_l.
\end{align}
Using this assumption, we get the integro-differential equation 
\begin{align}
\dfrac{d}{dt}a_{0}(t) =- \sum_{l=1}^L \dfrac{\gamma_l \lambda_l}{2}  \int_0^t dt' e^{-i \Delta_l (t-t')}e^{- \lambda_l(t-t')}  a_{0}(t'),
\label{a_0-3}
\end{align}
where $\Delta_l = \omega_l-\omega_0$.
From this equation, we can see that the parameter $\lambda_l$ represents how long past state affects the present dynamics.
If there exists finite $\lambda_l$, the present dynamics depends on the past dynamics.
Therefore, the system interacts with non-Markovian reservoir and its non-Markovianity is characterized by $\lambda_l$.

\begin{figure}[tb]
   \centering
   \includegraphics[width=0.9\columnwidth]{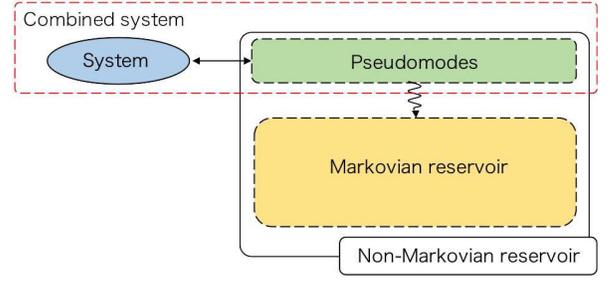}
\caption{Diagrammatic representation of the combined system dynamics. The system interacts with pseudomodes which leaks into the  Markovian reservoir. The interaction between the system and pseudomodes is effectively described by $H_{SP}$.and the leak rate form pseudomode $l$ into the Markovian reservoir is given by $2\lambda_l$.}
\label{system_image}
\end{figure}

With the pseudomode method \cite{PhysRevA.54.3592,PhysRevA.55.2290}, the dynamics of this system can be mapped into Markovian dynamics of a combined system of the two level system and $L$ pseudomodes system.
For the present model, pseudomode method leads to the following Markovian master equation
\begin{align}
\dfrac{d}{dt}\rho_{SP}^I(t) 
= \dfrac{1}{i \hbar}\left[H_{SP},\rho_{SP}^I(t)\right]
 + \sum_{l=1}^L 2 \lambda_l \mathcal{D}[c_l]\rho_{SP}^I(t),
 \label{meq}
\end{align}
where $\rho_{SP}^I(t)$ is the density operator of the combined system, $\mathcal{D}[\cdot]$ is a superoperator
\begin{align}
\mathcal{D}[A]\rho = A\rho A^\dag - \dfrac{1}{2}\left(A^\dag A \rho- \rho A^\dag A \right),
\end{align}
which describes the dissipation and $H_{SP}$ is a combined system Hamiltonian
\begin{align}
H_{SP} =  \sum_{l=1}^L \hbar\Delta_l c_l^\dag c_l + \sum_{l=1}^L \hbar\sqrt{\frac{\gamma_l\lambda_l}{2}} \left(\sigma_+ c_l+\sigma_-c_l^\dag\right),
\end{align}
where $c_l^\dag$ and $c_l$ are the creation and annihilation operators for the pseudomode labeled by $l$.
From the definition of the pseudomdoes, the initial state of the pseudomodes is the vacuum (see Refs.~\cite{PhysRevA.54.3592,PhysRevA.55.2290} for details). 
From Eq.~\eqref{meq}, we can see that the system coherently interacts with pseudomodes and each pseudomode dissipatively interacts with a Markovian reservoir (FIG. \ref{system_image}).
The information contained in the atom first flows to pseudomodes and then from each pseudomode to its reservoir.
The flow from each pseudomode to its reservoir is one-way, but the flow between the atom and pseudomodes is two-way.
In non-Markovian dynamics, the atom recieves back information and excitation energy from the reservoir due to memory effect.
Therefore pseudomodes could be seen as an effective description of the reservoir memory  \cite{PhysRevA.80.012104}.

To get the state of the system, we should trace out the pseudomodes,
\begin{align}
\rho_S^I(t) = {\rm Tr}_P\rho_{SP}^I(t).
\end{align}

\section{stochastic approach\protect\\ \label{SA}}
With the pseudomode method, the dynamics of this system is effectively described by the Markovian master equation.
So we can use stochastic approach for Markovian dynamics.
Here we define a non-Hermitian Hamiltonian, 
\begin{align}
H_{\rm eff}^I = H_{SP} - i\sum_{l=1}^L \hbar\lambda_l  c_l^\dag c_l .
\label{H_eff}
\end{align}
In this Hamiltonian, non-Hermitian term represents dissipation into the Marovian resevoir.
Using Eq.~\eqref{H_eff}, we can rewrite Eq.~\eqref{meq} as, 
\begin{align}
\dfrac{d}{dt}\rho_{SP}^I(t) 
=& \dfrac{1}{i \hbar}\left[H_{\rm eff}^I \rho_{SP}^I(t)-\rho_{SP}^I(t)(H_{\rm eff}^I)^\dag\right] \nonumber \\
& + \sum_{l=1}^L2\lambda_l c_l\rho_{SP}^I(t)c_l^\dag.
\end{align}
The first term of the right-hand side represents the continuous dynamics governed by the non-Hermitian Hamiltonian $H_{\rm eff}^I$.
The second term represents jump process which is the loss of an excitation energy from pseudomodes.

Using an unnormalized state vector $|\tilde{\Psi}(t)\r$ which satisfies Schr\"{o}dinger equation,
\begin{align}
i \hbar \dfrac{d}{dt}|\tilde{\Psi}(t)\r = H_{\rm eff}^I|\tilde{\Psi}(t)\r,
\label{sch-eq}
\end{align}
we can divide $\rho_{SP}^I(t)$ into two terms as follows,
\begin{align}
\rho_{SP}^I(t) = |\tilde{\Psi}(t)\r\l \tilde{\Psi}(t)|+ \Pi_p(t) |g,0_P\r\l g,0_P|,
\label{Psi_SP}
\end{align}
where $|0_P\r$ is the vacuum state of pseudomodes.
The trace of $\rho_{SP}^I(t)$ is conserved to 1, so that the coefficient of the second term $\Pi_p(t)$ is defined by $\Pi_p(t)= 1 - \l\tilde{\Psi}(t)|\tilde{\Psi}(t)\r$ of time.
Because $\lambda_l$ is defined as positive, the inner product of $|\tilde{\Psi}(t)\r$ is a monotonic decreasing function of time and $\Pi_p(t)$ is a monotonic increasing function.

From the quantum trajectory approach \cite{wiseman2010quantum}, the unnormalized state vector $|\tilde{\Psi}(t)\r$ is a trajectory under no jumps.  
Since the system is two level atom and the initial state of the reservoir is vacuum, the jumped part (= the second term of Eq.~\eqref{Psi_SP}) is the ground state.
The state of the two level atom is given by 
\begin{align}
\rho_S^I(t) = {\rm Tr}_P |\tilde{\Psi}(t)\r\l \tilde{\Psi}(t)| + \Pi_p(t) |g\r \l g|.
\end{align}

The probability that there is no jump until time $t$ (= the survival probability) is given by the inner product of $|\tilde{\Psi}(t)\r$,
\begin{align}
P_0(t) = \l\tilde{\Psi}(t)|\tilde{\Psi}(t)\r.
\end{align}
Because we can regard the pseudomodes as memory part of the reservoir \cite{PhysRevA.80.012104},
the survival probability $P_0(t)$ can be regarded as the probability that the system interacts with its reservoir coherently until time $t$.
The jump rate to the ground state of the combined system, which is given by the damp rate of $P_0(t)$, represents a memory loss rate.
So the probability density of jump is given by
\begin{align}
p(t) = -\dfrac{d}{dt}  P_0(t).
\label{p_t}
\end{align}
This probability density represents the information flux from pseudomode into Markovian reservoir.
For the particular model, the relationship between the oscillation of $p(t)$ and the measure of non-Markovianity had been discussed \cite{PhysRevA.90.054101}.

Since the initial state of the atom is $|\psi(0)\r = \alpha |e\r + \beta |g\r$,
the state of the combined system can be expanded as
\begin{align}
|\tilde{\Psi}(t)\r= \alpha(a_{0}(t)|e,0_P\r+\sum_{l=1}^L q_l(t)|g,1_l\r)+\beta|g, 0_P\r, 
\label{Psi_l}
\end{align}
where $|1_l\r = c_l^\dag |0_P\r$, $a_0(0)=1$ and $q_l(0)=0$.
This $a_0(t)$ is the same as the amplitude $a_0(t)$ in Eq.~\eqref{a_0-3}.
Using Eq.~\eqref{Psi_l}, the survival probability $P_0(t)$ and the probability density $p(t)$ are given by
\begin{align}
P_0(t) = | \alpha |^2\left( |a_{0}(t)|^2 +\sum _{l=1}^L|q_l(t)|^2\right) +|\beta|^2,
\end{align}
\begin{align}
p(t) = \sum_{l=1}^L 2\lambda_l|\alpha q_l(t)|^2.
\end{align}
The probability density $p(t)$ is the $2\lambda_l|\alpha q_l(t)|^2$ represents the energy flow from pseudomode $l$ to its reservoir.

If the jump to the ground state occurs during the measurement time $T$, the expectation value of the jump time is given by
\begin{align}
\left\l t \right\r_T  = \dfrac{\disp\int_0^T t p(t) dt}{\disp\int_0^T p(t)dt},
\end{align}
and we define the expectation value $\left\l t \right\r $ as the long measurement time limit of $\left\l t \right\r_T $,
\begin{align}
\left\l t \right\r   \equiv& \lim_{T\to \infty} \left\l t \right\r_T \nonumber \\
=& \int_0^\infty (| a_{0}(t)|^2 +\sum _{l=1}^L| q_l(t)|^2)dt.
\end{align}
Moreover we define that
\begin{align}
 \left\l t_S \right\r=& \int_0^\infty | a_{0}(t)|^2dt , \label{t_S_ave}\\
\left\l t_l \right\r =&  \int_0^\infty | q_l(t)|^2dt, \label{t_l_ave}
\end{align}
and then we get $\left\l t \right\r =  \left\l t_S \right\r + \sum_l \left\l t_l \right\r$,
where $ \left\l t_S \right\r$ is the expected time length that the two level system is in the excited state $|e\r$ and $\left\l t_l \right\r$ is one that a pseudomode $l$ is in the excited state $|1_l\r$.
Since we can regard the pseudmodes as the degree of freedom of the reservoir that interact with the system of interest coherently \cite{PhysRevA.80.012104},
$\sum_l \left\l t_l \right\r$ can be regarded as the expectation value of memory time of the non-Markovian reservoir and reflects non-Markovianity of the system dynamics.

We consider the Markovian limit ($\lambda_l\to\infty$).
When $\lambda_l \gg \Delta_l ,\gamma_l$, the unnormalized state vector is approximated as 
\begin{align}
|\tilde{\Psi}(t)\r \simeq \left(\alpha e^{-\frac{1}{2}\sum_l \gamma_l t} |e\r +\beta |g\r \right) |0_P\r.
\end{align}
Therefore, we can see that $q_l(t)=0$ for any $t>0$ and $\left\l t_l \right\r = 0$ in the Markovian limit .
Because time $t$ is positive, $\left\l t_l \right\r=0$ means that there is no time that pseudomodes are in their excited states.
The Markovian limit is the limit the reservoir has no memory.
This is consistent and intuitive with the result we have got here; pseudomodes are vanished and $\sum_l \left\l t_l \right\r$ converges to $0$ in the Markovian limit, so that pseudomode is a memory part of the reservoir and $\sum_l \left\l t_l \right\r$ is an expectation value of memory time of the reservoir.
This result also suggests the following criterion.
\begin{itemize}
\item
When $\sum_l \l t_l \r = 0$, its dynamics is Markovian.
\item
When $\sum_l \l t_l \r \neq 0$, its dynamics is non-Markovian.
\end{itemize}

\section{Damped Jaynes-Cummings model\protect\\ \label{DJC}}
In this section, we discuss the dynamics of a two level atom in a lossy cavity \cite{breuer2007theory}.
The reservoir is electromagnetic field inside and outside the cavity and its density of state has a peak at the cavity resonant frequency.
Therefore, we can assume that the structure is a single Lorentz function,
\begin{align}
D(\omega) = \dfrac{\gamma \lambda^2}{(\omega -\omega_c)^2 + \lambda^2},
\end{align}
where $\omega_c$ is the resonant frequency of the cavity.
This is called the damped Jaynes-Cummings model,  which is a typical example of non-Markovian system and $L=1$ case of what we have discussed.
Therefore, we can use pseudomode method and the result we have got.
The effective Hamiltonian $H_{\rm eff}^I $ is
\begin{align}
H_{\rm eff}^I &= H_{SP} - i\hbar\lambda c_p^\dag c_p \nonumber \\
&=  \hbar\left(\Delta- i \lambda \right) c_p^\dag c_p +\hbar\sqrt{\dfrac{\gamma\lambda}{2}}\left(\sigma_+ c_p+\sigma_-c_p^\dag\right),
\end{align}
and the unnormalized state $|\tilde{\Psi}(t)\r$ is
\begin{align}
|\tilde{\Psi}(t)\r= \alpha(a_{0}(t)|e, 0_P\r+ q(t)|g,1_P\r)+\beta|g, 0_P\r,
\end{align}
where $\Delta=\omega_c-\omega_0$ is the detuning between the two level system and  the pseudomode, $c_p^\dag$ and $c_p$ are creation and annihilation operators for the pseudomode and $|1_P\r = c_p^\dag |0_P\r$.

\begin{figure*}[htbp]
\centering
\subfigure{\includegraphics[width=0.9\columnwidth]{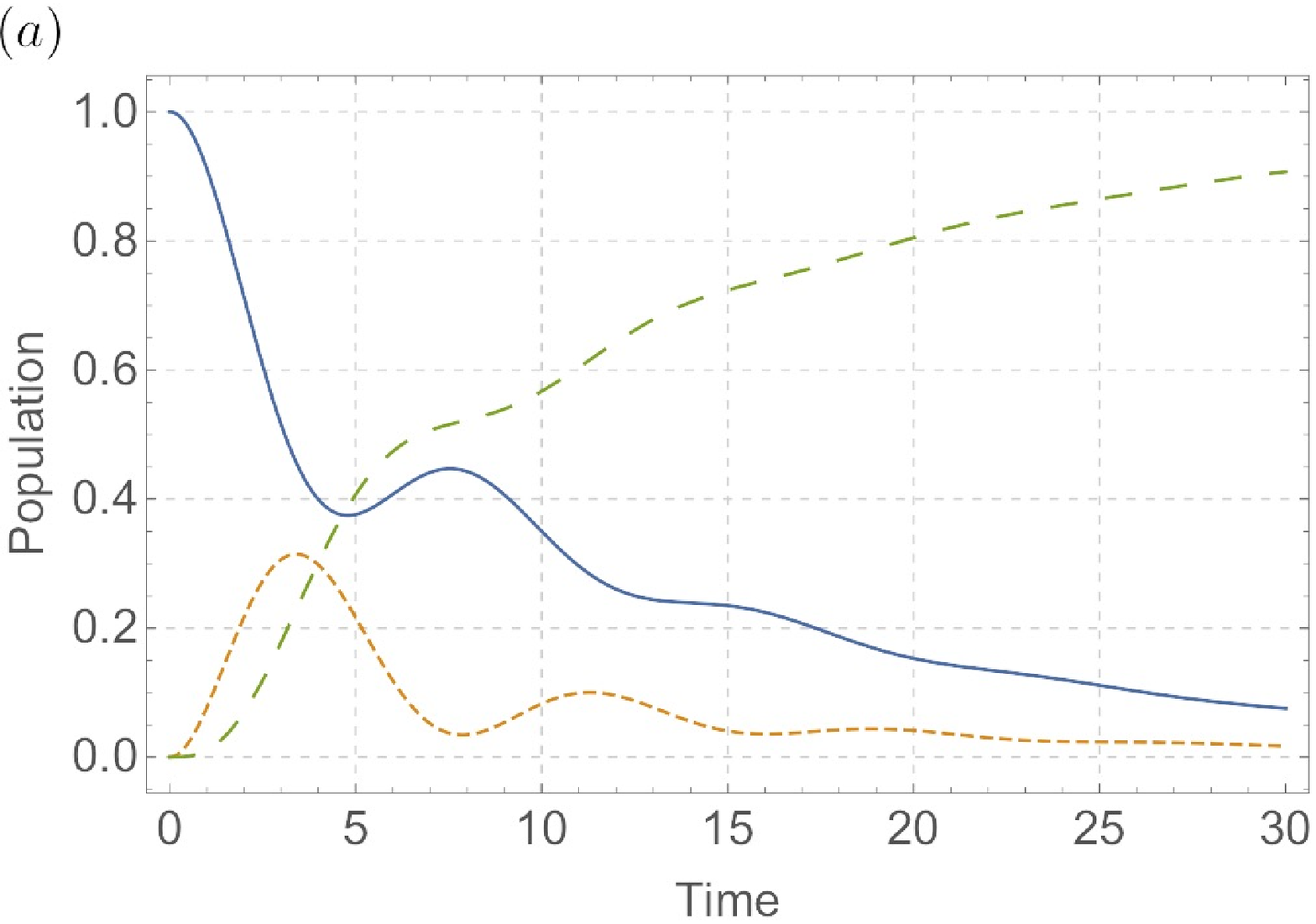}\label{a0_q}}
\qquad\qquad
 \subfigure{\includegraphics[width=0.9\columnwidth]{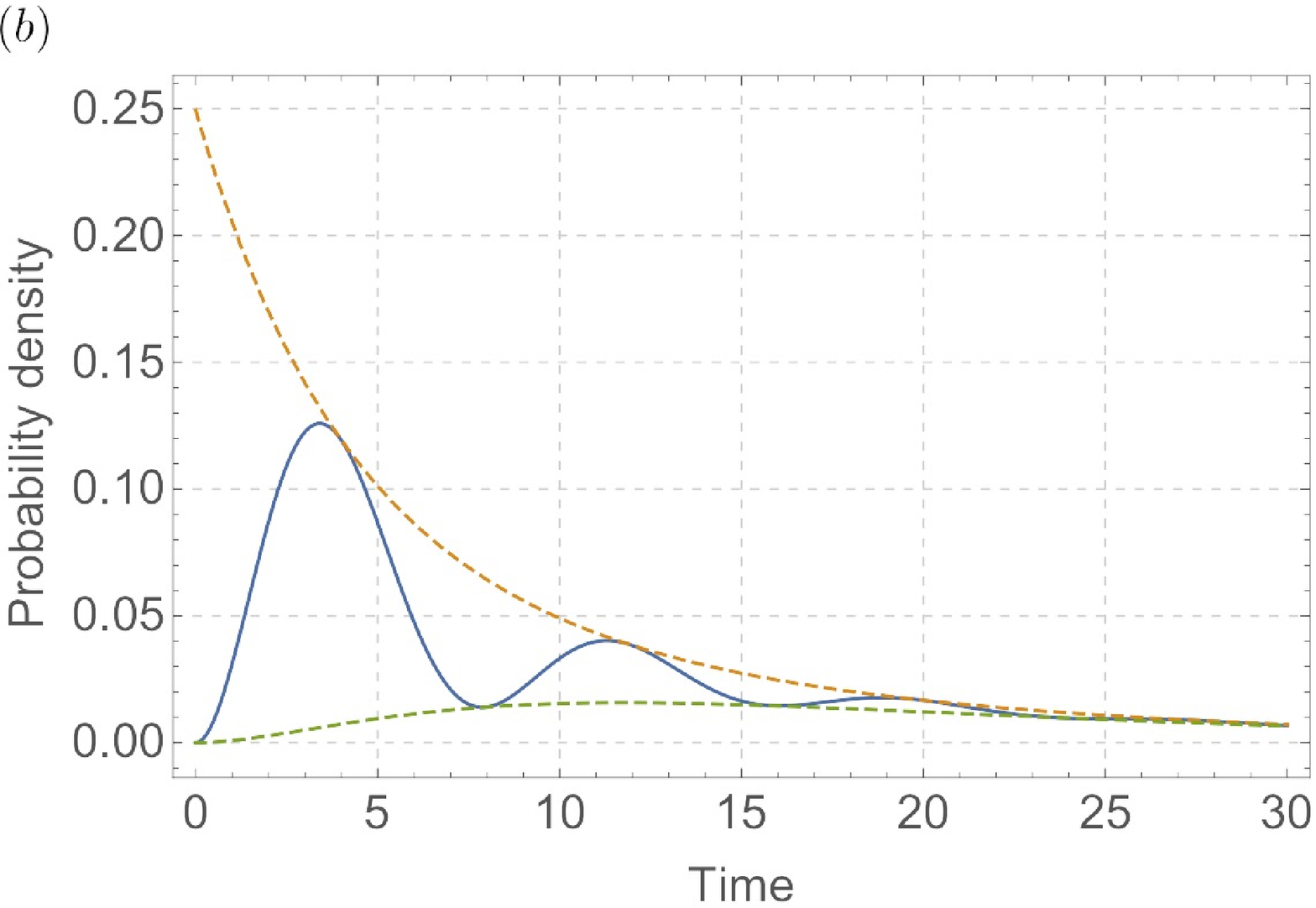}\label{p_detuned_oc}}
  \caption{(Color online) Population and probability density. The initial condition is exited state ($\alpha=1$, $\beta=0$).
The parameters are $\gamma=1$, $\lambda=0.2$ and $\Delta=0.5$. With these parameters, $\left\l t_S \right\r = 9.8$ and $\left\l t _P \right\r=2.5$. (a) Populations $|a_{0}(t)|^2$ (blue solid), $|q(t)|^2$ (yellow dashed) and $\Pi_p(t)$ (green large dashed). Because of the definition, these populations satisfy $|a_{0}|^2+|q(t)|^2+\Pi_p(t)=1$.  (b) Probability density $p(t)$ (blue solid). Yellow and green dashed lines are the plots in which we replace oscillation term with $\pm1$.}
\end{figure*}

Inserting the effective Hamiltonian $H_{\rm eff}^I $ and  the unnormalized state $|\tilde{\Psi}(t)\r$ into Schr\"{o}dinger equation, we get two simultaneous differential equations,
\begin{align}
\left\{
\begin{array}{l}
i\dfrac{d}{dt}a_{0}(t) =\sqrt{\dfrac{\gamma \lambda}{2}}q(t) \\
\ i\dfrac{d}{dt}q(t) = (- i\lambda+\Delta) q(t) +\sqrt{\dfrac{\gamma \lambda}{2}}a_{0}(t).\\
\end{array}
\right.
\end{align}
Eigenvalues of these equations are
\begin{align}
\dfrac{-(\lambda + i \Delta) \pm \sqrt{(\lambda + i \Delta)^2 -2\gamma \lambda}}{2} .
\end{align}
Under the initial condition, $a_{0}(0)=1$ and $q(0)=0$, 
the solution is
\begin{align}
\left\{
\begin{array}{l}
a_{0}(t) = e^{-\frac{\lambda}{2}t}e^{-i\frac{\Delta }{2}t}\left[\cosh\left(\dfrac{dt}{2}\right)+\dfrac{\lambda+i\Delta}{d}\sinh\left(\dfrac{dt}{2}\right)\right], \\
\ q(t)  = - i \frac{\sqrt{2\gamma\lambda}}{d}  e^{-\frac{\lambda}{2}t} e^{-i\frac{\Delta}{2}t}\sinh\left(\dfrac{dt}{2}\right) ,
\end{array}
\right.
\label{result_tau}
\end{align}
where we define $d =\sqrt{ (\lambda + i \Delta)^2 - 2\gamma \lambda}$.
As a result, we get the probability density,
\begin{align}
p(t) =& 2\lambda |\alpha q(t)|^2 \nonumber \\
=& \dfrac{2 |\alpha|^2 \gamma\lambda ^2}{|d|^2} e^{-\lambda t} \left(\cosh({\rm Re}[d] t)-(\cos({\rm Im}[d] t)\right) .
\label{p_t_DJCM}
\end{align}

FIG.~\ref{a0_q} shows the dynamics of the populations as a function of time. 
In FIG.~\ref{a0_q}, $|a_0(t)|^2$ and $|q(t)|^2$ oscillate and $\Pi_p(t)$ monotonically increases.
The correlated oscillations of $|a_0(t)|^2$ and $|q(t)|^2$ are caused by non-Markovianity.
There is no energy flow from Markovian reservoir into the combined system, so that $\Pi_p(t)$ monotonically increases.
FIG.~\ref{p_detuned_oc} shows the dynamics of the probability density of jump as a function of time. 
From FIG.~\ref{p_detuned_oc}, we can see that $p(t)$ is positive except for $t=0$ and $t\to\infty$, 
because $\cosh x >1$ for $\forall x>0$.
When there is no detuning $\Delta=0$, $d=\sqrt{\lambda^2-2\gamma\lambda}$, so that $d$ is real or pure imaginary.
When $2\gamma<\lambda$, $d$ is real so that $p(t)\neq 0$ for $t>0$.
On the other hand, when $\lambda<2\gamma$, $d$ is pure imaginary so that
\begin{align}
p(t) = \dfrac{ 2|\alpha|^2\gamma\lambda}{2\gamma-\lambda} e^{-\lambda t} \left(1-\cos\left(\sqrt{2\gamma\lambda-\lambda^2} t\right)\right).
\end{align}
Therefore when the time satisfies
$
t= 2\pi  n(2\gamma-\lambda)^{-\frac{1}{2}}
$
for $n \in \{0,\mathbf{N}\}$, the pribability density $p(t)$ is $0$.

Here the structure of the reservoir is a single Lorentz function so that there is only a single pseudomode and the expectation value of jump time is given by
\begin{align}
\left\l t \right\r \ =& \left\l t_S \right\r + \left\l t_P \right\r, \\
& \left\l t_S \right\r = \int_0^\infty | a_{0}(t)|^2dt,  \\
&\left\l t_P \right\r =  \int_0^\infty | q(t)|^2dt.
\end{align}
Using the result we calculated above, we get
\begin{align}
\left\{
\begin{array}{l}
\left\l t_S \right\r  = \dfrac{1}{\gamma}\left(1+ \left(\dfrac{\Delta}{\lambda} \right)^2\right)+ \dfrac{1}{2\lambda} \\
\left\l t_P \right\r  = \dfrac{1}{2\lambda}.
\end{array}
\right.
\label{t_sp}
\end{align}

\begin{figure*}[htbp]
\centering
\subfigure{\includegraphics[width=0.9\columnwidth]{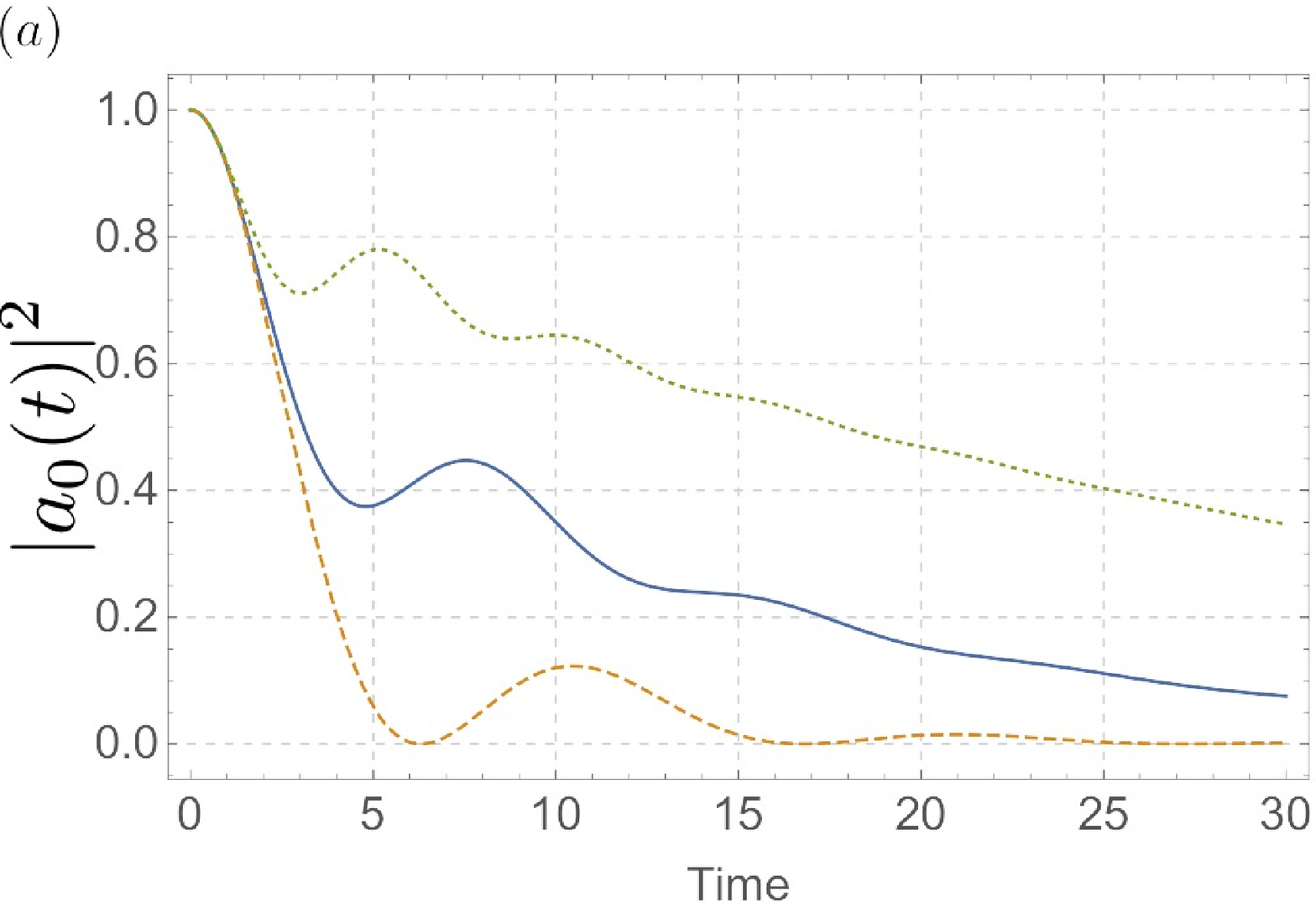}\label{a0_detuning}}
\qquad\qquad
 \subfigure{\includegraphics[width=0.9\columnwidth]{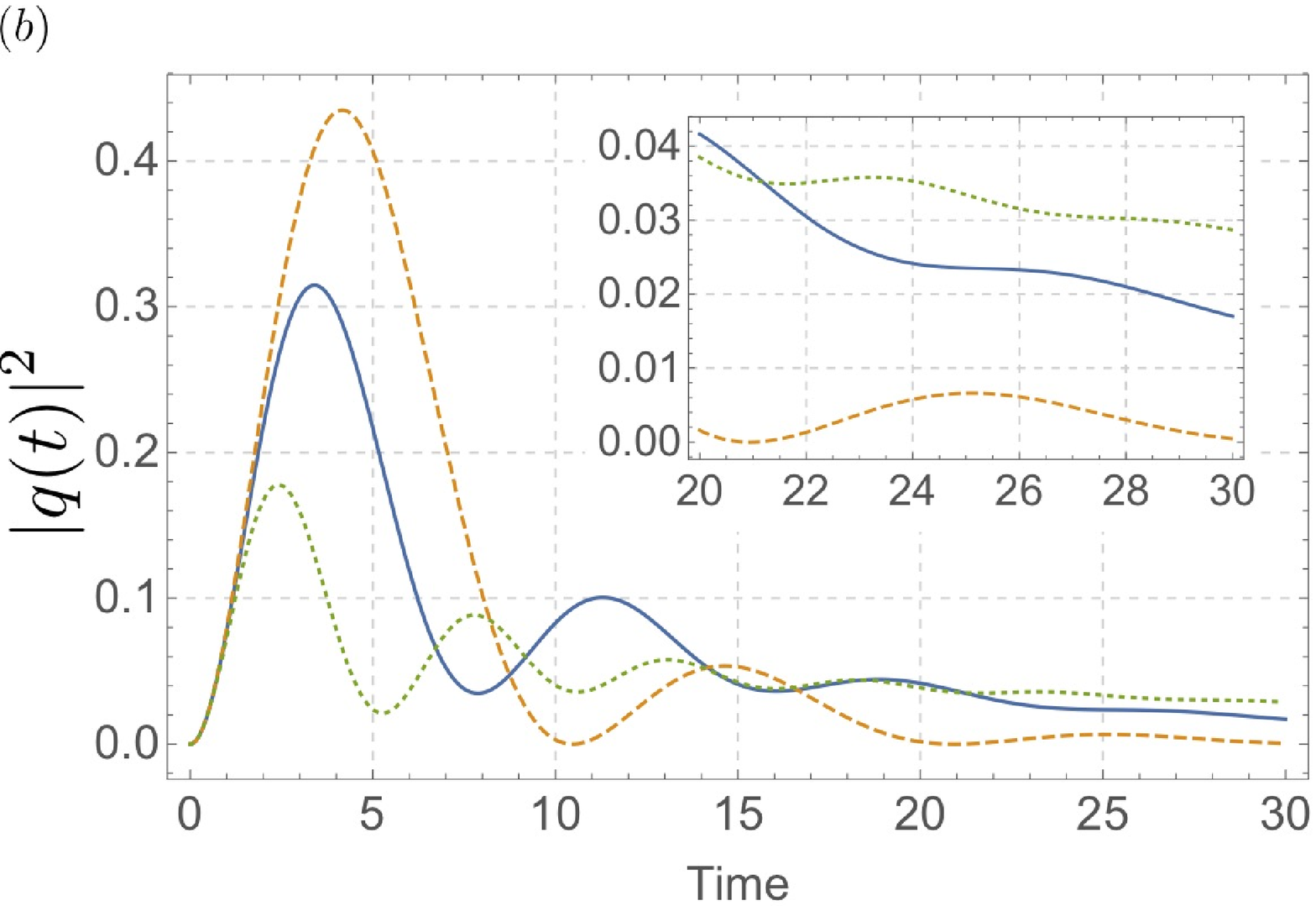}\label{q_detuning}}
\caption{ (Color online) (a) The system excited state population $|a_0(t)|^2$ and (b) Pseudomode excited state population $|a_p(t)|^2$ with some detuning values; Blue solid: $\Delta=0.5$, Yellow dashed: $\Delta =0$ and Green dotted: $\Delta=1$. The other parameters are same ($\gamma=1$, $\lambda=0.2$). (a) The larger detuning makes the damping more slowly. (b) When the detuning $\Delta$ increases, the maximum value of $|q(t)|^2$ decreases and the excited state population increases in the later time (see the inset). }
\label{a0_q_detunig}
\end{figure*}

As noted at Eq.~\eqref{a_0-3}, non-Markovianity is characterized by $\lambda$.
The decrease of $\lambda$ means an increase of the reservoir correlation time, hence the non-Markovianity becomes stronger.
In Eq.~\eqref{t_sp}, the expectation values $\l t_S \r$ and $\l t_P\r$ are monotonically decreasing functions for  $\lambda$.
Therefore, the result shows that non-Markovianity of the system dynamics is reflected to the delay of the expectation value of the jump time $\left\l t \right\r$.

From Eq.~\eqref{t_sp}, we see that $\l t_S \r$ depends on the detuning $\Delta$ and $\l t_P\r$ do not depend on it.
This can be understood as follows.
In the detuned Rabi oscillation, the oscillating amplitude is smaller than unity, which depends on the value of the detuning.
The damp rate of the system excited state population and the maximum value of the pseudomode excited state population are suppressed by increasing the detuning, which are shown in FIG.~\ref{a0_q_detunig}.
Therefore, the damp of $P(t)$ is slower than resonance case and the expected time length that the atom is in the exited state increases as the detuning increases.
However, the leak rate from the pseudomode into the Markovian reservoir is $2\lambda$, which does not depend on the detuning.
This is the reason why the expected time length the pseudomode is in the exited state is the invariant value for the detuning. 
As shown in FIG.~\ref{q_detuning}, instead of the suppression of the maximum value, the population in the later time increases.

We  also calculate the generating function
\begin{align}
\chi(\omega) \equiv  \int_0^\infty p(t) e^{i\omega t} dt.
\end{align}
Using Eq.~\eqref{result_tau}, we get explicit form of the generation function
\begin{align}
\chi(\omega) = \dfrac{2\gamma\lambda^2(\lambda-i\omega)}{(\lambda-i\omega)^4-(\lambda^2-\Delta^2-2\gamma\lambda)(\lambda-i\omega)^2-(\Delta\lambda)^2}.
\end{align}
From this function, we can get the expectation value
\begin{align}
\l t \r =& \left. \dfrac{d\ln \chi(\omega)}{d (i\omega)}\right|_{\omega=0}\nonumber  \\
=&  \dfrac{1}{\gamma}\left(1+\left(\dfrac{\Delta}{\lambda}\right)^2\right)+\dfrac{1}{\lambda},
\end{align}
and the variance
\begin{align}
\l (\delta t)^2 \r 
=& \left. \dfrac{d^2\ln \chi(\omega)}{d (i\omega)^2}\right|_{\omega=0} \nonumber \\
=& \dfrac{1}{\gamma^2}\left(1+\left(\dfrac{\Delta}{\lambda}\right)^2\right)^2-\dfrac{1}{\gamma\lambda}\left(1-3\left(\dfrac{\Delta}{\lambda}\right)^2\right)+\dfrac{1}{\lambda^2}. \nonumber \\
\end{align}
In order to understand the relationship between these values, we define the function
\begin{align}
\Lambda_{\l t\r} \equiv& \dfrac{\l (\delta t)^2 \r-(\l t \r )^2}{(\l t \r )^2} \nonumber \\
=& -\dfrac{(3\lambda^2-\Delta^2)\gamma\lambda}{(\lambda^2+\gamma\lambda+\Delta^2)^2}.
\end{align}
This function can be divided into 3 cases as follows
\begin{align}
\Lambda_{\l t\r}= 
\left\{
\begin{array}{lcl}
 >0 &\cdots& \sqrt{3}\lambda < |\Delta| \\
=0 &\cdots& \sqrt{3}\lambda = |\Delta| \\
<0 &\cdots& \sqrt{3}\lambda > |\Delta| 
\end{array}
\right.
\end{align}
The sign of $\Lambda_{\l t\r}$ changes at $\lambda_0=|\Delta|/\sqrt{3}$.
In the Markovian limit ($\lambda\to \infty$), the expectation value and the variance converge to $\l t \r \to \gamma^{-1}$ and $\l (\delta t)^2 \r \to \gamma^{-2}$, respectivelity.
Thus $\Lambda_{\l t\r}$ converges to $0$ in the Markovian limit.
Because the correlation time $\lambda_0^{-1}=\sqrt{3}/|\Delta|$ is small for large detunig $|\Delta|$,  the function $\Lambda_{\l t\r}$ is positive for relatively large $\lambda$ for large detuning.
When it is negative,  the variance is relatively smaller than one of  Markovian dynamics.
 $\left\l t \right\r$ is the expected time length that the system and the reservoir can interact with each other coherently.
 Therefore, negative $\Lambda_{\l t\r}$ means that the memory is lost at more definite time compared with Markovian dynamics.

\section{Conclutions\protect\\ \label{conclude}}
We have studied the non-Markovian dynamics of a two level atom, using pseudomode method and the stochastic approach for Markovian dynamics.
In this paper, we have assumed that the structure of the reservoir is given by a sum of Lorentz functions.
With pseudomode method, the non-Markovian dynamics of a two level atom can be mapped to Markovian dynamics of a combined system of the system and pseudomodes whose number is the same as that of the Lorentz functions.

The expectation value of jump time to the ground state of the combined system is given by the sum of the expected time length that the two level system is in the exited state and one that each pseudomode is in its excited state.
The later time length represents the memory time of a non-Markovian reservoir.
In the Markovian limit, we get the result that the probability that pseudomodes are in their exited states is 0.
Then the expected time length that pseudomodes in their excited state also converges to 0.

In particular, we have discussed the damped Jaynes-Cummings model, which is a model of a two level atom in a lossy cavity.
This is analytically solvable so that we can get an exact solution of the dynamics and the expectation values, explicitly.
As a result, we have found that the expected time length that the system and the pseudomode are in their excited state took the value reflecting non-Markovianity.

Since Markovian approximation is the approximation that the reservoir has no memory, 
we can say that our result suggest the pseudomode is the degree of the freedom characterizing the memory of the reservoir. 

\section*{ackowledgements}
The authors are grateful to C. Uchiyama for thoughtful comments and suggestions.
This work was supported by CREST, JST.

\bibliography{MemoryEffect_PM}

\end{document}